\documentclass{ptephy_v1}\usepackage[version=4]{mhchem}
\usepackage{multirow}
\usepackage{gensymb}
\usepackage{url}
\usepackage{booktabs}
\usepackage{pbox}

\preprintnumber{XXXX-XXXX}

\begin{document}

\title{Study of radon removal performance of silver-ion exchanged zeolite from air for underground experiments}

\author{T. Sone}
\affil{Department of Physics, Graduate School of Science, Kobe University, Kobe, Hyogo 657-8501, Japan}

\author[1,2,*]{Y. Takeuchi}
\affil{Kavli Institute for the Physics and Mathematics of the Universe~(WPI), The University of Tokyo Institutes for
Advanced Study, University of Tokyo, Kashiwa, Chiba 277-8583, Japan}

\author{M. Matsukura}
\affil{Institute of Engineering Innovation, School of Engineering, The University of Tokyo, Bunkyo, Tokyo 113-8656, Japan}

\author{Y. Nakano}
\affil{Faculty of Science, University of Toyama, Toyama 930-8555, Japan}

\author{H. Ogawa}
\affil{CST Nihon University, Chiyoda, Tokyo 180-0011, Japan}

\author[6,2]{H. Sekiya}
\affil{Kamioka Observatory, Institute for Cosmic Ray Research, the University of Tokyo, Gifu 506-1205, Japan}

\author[7,3]{T. Wakihara}
\affil{Department of Chemical System Engineering, The University of Tokyo, Bunkyo, Tokyo 113-8656, Japan}

\author{S. Hirano}
\affil{Tosoh Corporation, Chuo, Tokyo 104-8467, Japan}

\author{A. Taniguchi}
\affil{Sinanen Zeomic Co., Ltd., Nagoya, Aichi 455-0051, Japan\email{takeuchi@phys.sci.kobe-u.ac.jp}}

\begin{abstract}
Radon is a common background source for underground astroparticle physics experiments and its removal is essential.
In this study, we fabricated several prototype silver-ion exchanged zeolites,
measured their radon adsorption properties, and evaluated their applicability for radon removal
in underground experiments.
As a result, we succeeded in producing a prototype silver-ion exchanged zeolite that showed better radon adsorption
properties than preceding studies.
 
\end{abstract}

\subjectindex{H20}

\maketitle

\section{Introduction}

Since underground location can greatly reduce the effects of cosmic rays, it is ideal for the astroparticle physics
experiments that require extremely low background, such as solar and supernova neutrino observation experiments and
direct dark matter search experiments (hereafter referred to as underground experiments).
One of the main backgrounds common to underground experiments is the radioactive noble gas radon-222
(hereafter referred to as radon, Rn).
For example, the radon concentration in the ambient air near the Super-Kamiokande detector, the world's largest
neutrino detector located underground in Kamioka, Japan, ranges from 50 to 2000 Bq/m$^3$~\cite{SK-radon,SK-radon2},
but for accurate solar neutrino observation, the radon concentration in water in the detector must be kept below
around 1 mBq/m$^3$.
In the XENONnT experiment, which is a direct search for dark matter in Gran Sasso, Italy,
the radon concentration in liquid xenon must be kept below about 4 $\mu$Bq/kg~\cite{XENON1}.
Radon removal by activated carbon columns~\cite{Fukuda:2002uc,Nakano:2017rsy} and radon removal
by distillation equipment~\cite{XENON2} have been developed and operated for reducing radon
in these experiments, but both are rather large devices.
Due to the high space cost in the underground experimental area, the future planned experiments
will require smaller equipment for radon removal.

On the other hand, in industry, zeolite, a crystalline porous material with homogeneous pores of about 0.2 to 1.0 nm,
is widely used as an adsorbent, desiccant, ion exchanger, catalyst, wastewater treatment, and so on.
Silver-ion exchanged zeolite (Ag-zeolite), in which silver ions are introduced into the zeolite by ion exchange,
has also been commercialized as a material with antibacterial and deodorizing properties.
There are various types and applications of Ag-zeolites, and various research reports have been done,
for example, at the recent conferences held by the Japan Zeolite Association, and/or the Japan Society on Adsorption.
An example of a research topic related to this study is the report that Ag-FER (Ferrierite) exhibit
excellent separation and adsorption properties for xenon in the atmosphere~\cite{Fukui}.

In addition, recent studies have reported that Ag-zeolites have very high adsorption
performance with respect to radon in air~\cite{zeo1,zeo2}.
It is reported that the radon adsorption performance of Ag-ETS-10 (Engelhard Titano-Silicate, a zeolite analog)
and Ag-ZSM-5 (zeolite soconi Mobil-FIve, or Ag-MFI)
at room temperature is equivalent to about 500 times that of activated carbon, which could
significantly reduce the size of conventional large-scale radon removal systems.
Our research group also produced Ag-FER and Ag-MFI in powder form in 2023 and confirmed their ability
to adsorb radon in air at room temperature~\cite{ogawa2}. 
However, it is also reported in these papers~\cite{zeo1,zeo2} that the adsorption capacity of radon is significantly
reduced in high-humidity air.
While radon adsorption by granular activated carbon and activated carbon fiber uses physical adsorption
due to van-del-Waals forces in the pores,
it has been pointed out that the strong adsorption performance of xenon and other noble gases
by Ag-zeolite is due to chemical adsorption by silver (metallic silver or silver ions),
which has not yet been understood at the molecular level~\cite{zeo3}.
In any case, if chemical adsorption for radon is possible, radon removal should be much more powerful
than the traditional radon removal using physical adsorption in the underground experimental field.

The activated carbons used in underground experiments have reached a level of intrinsic radon emanation 
in the order of less than 10 mBq/kg~\cite{ac1,ac2}.
In order to apply Ag-zeolites in underground experiments, the intrinsic radon emanation
from the Ag-zeolite should be kept below the similar level.
Industrial products are generally not concerned with radon emanation amounts,
and the studies in Ref.~\cite{zeo1} were done at ~kBq/m$^3$ radon concentration levels.
Ref.~\cite{zeo2} reported a radon emanation of $71 \pm 1$ mBq/kg from Ag-ETS-10,
and Ref.~\cite{ogawa} reported a radon emanation of $14.2 \pm 7.0$ mBq/kg from 
5A zeolite (Ca-Zeolite A) synthesized using low-radioactivity raw materials.

In this study, we focused on the Ag-FER, which was reported to show excellent xenon adsorption performance
in air in Ref.~\cite{Fukui} and our research group also confirmed its radon adsorption ability in air.
We produced new Ag-zeolite samples and measured the radon concentration in air after processing with the Ag-zeolite,
the humidity dependence of the radon adsorption performance in air,
and the silver introduction dependence of the radon adsorption performance.
The results of these measurements are reported in this article.

\section{Preparation of Ag-FER zeolites}

In this study, samples were prepared from a pellet-shaped ferrielite zeolite HSZ-722HOD1A(H-type)
manufactured by Tosoh Corporation, and doped with silver at Sinanen Zeomic Co., Ltd.
Table~\ref{table_zeo1} shows the basic properties of the raw material zeolite.
\begin{table}[h]
\caption{Summary of basic properties of raw zeolite made by Tosoh Corporation.}
\label{table_zeo1}
\centering
\begin{tabular}{lcccccc}    \toprule
Crystal    & Pore & Shape & Binder & Cation & SiO$_2$/Al$_2$O$_3$  & Bulk    \\
form       & size &       & type   &        & ratio                & Density \\ \midrule
Ferrierite & 4.8 \AA & 1.5 mm$\phi$ pellet & Alumina & H$^+$ & 18 mol/mol & 0.56 Kg/L \\ \bottomrule
\end{tabular}
\end{table}
The raw zeolite is in the form of 1.5 mm diameter pellets solidified with alumina binder and has
a nominal pore size of 4.8 Angstroms.
Table~\ref{table_zeo2} shows the properties of the Ag-zeolite produced in this study.
\begin{table}[h]
\caption{Summary of basic properties of silver zeolite made by Sinanen Zeomic Co., Ltd.}
\label{table_zeo2}
\centering
\begin{tabular}{lccc}    \toprule
          & Target Silver & Silver introduction process & Measured Silver \\
          & Amount        &                        & Amount          \\ \midrule
3Ag-FER   & 3\%           & Normal ion exchange    & 3.2\%           \\
8Ag-FER-D & 8\%           & Impregnation and drying  & 8.1\%           \\
8Ag-FER-B & 8\%           & Special ion exchange   & 8.4\%           \\ \bottomrule
\end{tabular}
\end{table}
In this study, samples with silver contents of about 3\% and about 8\% were produced.
The 3\% sample was produced by the conventional ion-exchange method, while the 8\% samples
were produced using two different silver introduction methods.

\begin{figure}[hbt]
\centering\includegraphics[width=3.3in]{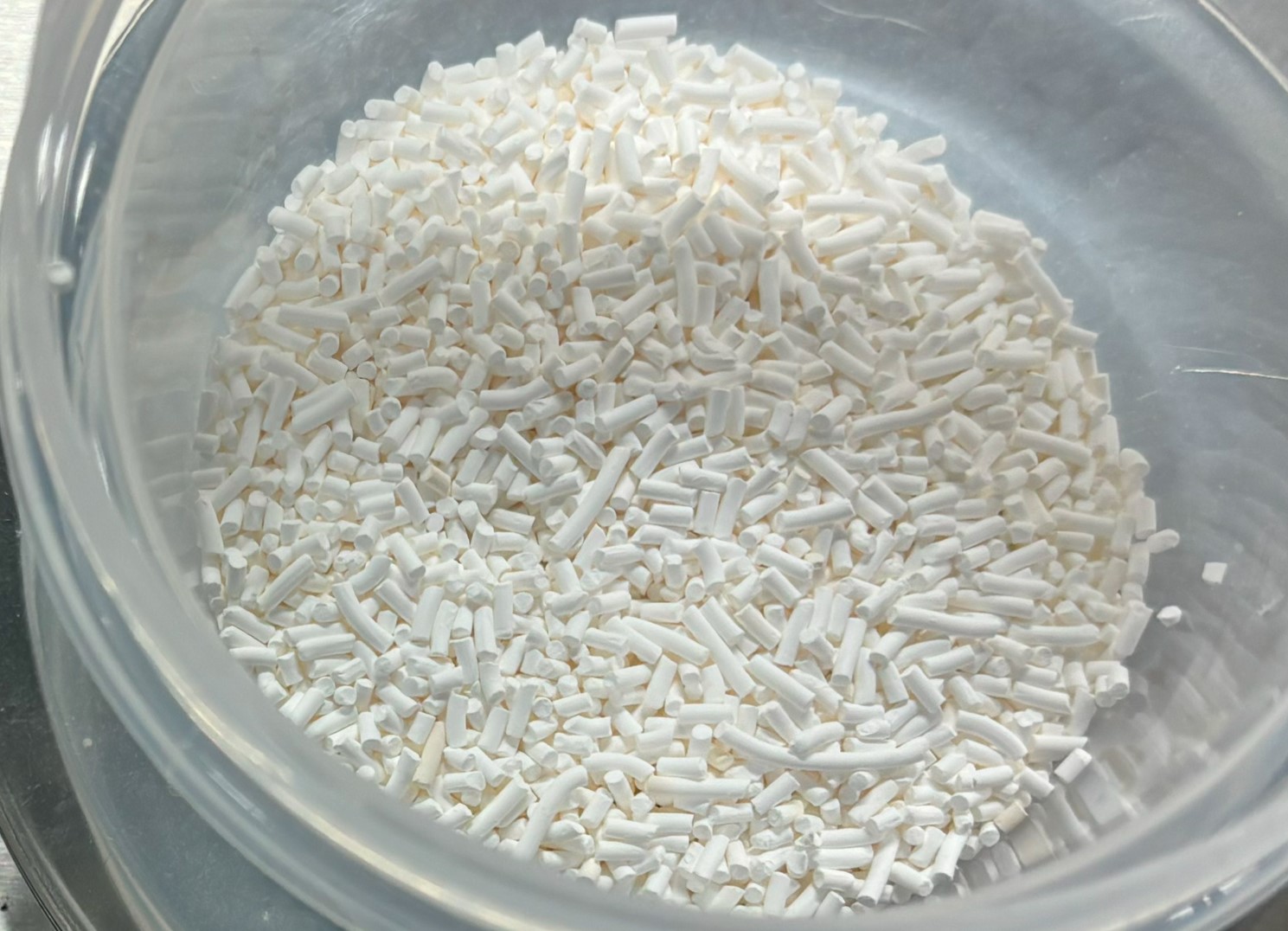}
\caption{Ag-zeolite pellet (8Ag-FER-D) made by Tosoh Corporation and Sinanen Zeomic Co., Ltd.}
\label{fig_zeo1}
\end{figure}

\begin{figure}[hbt]
\centering\includegraphics[width=5in]{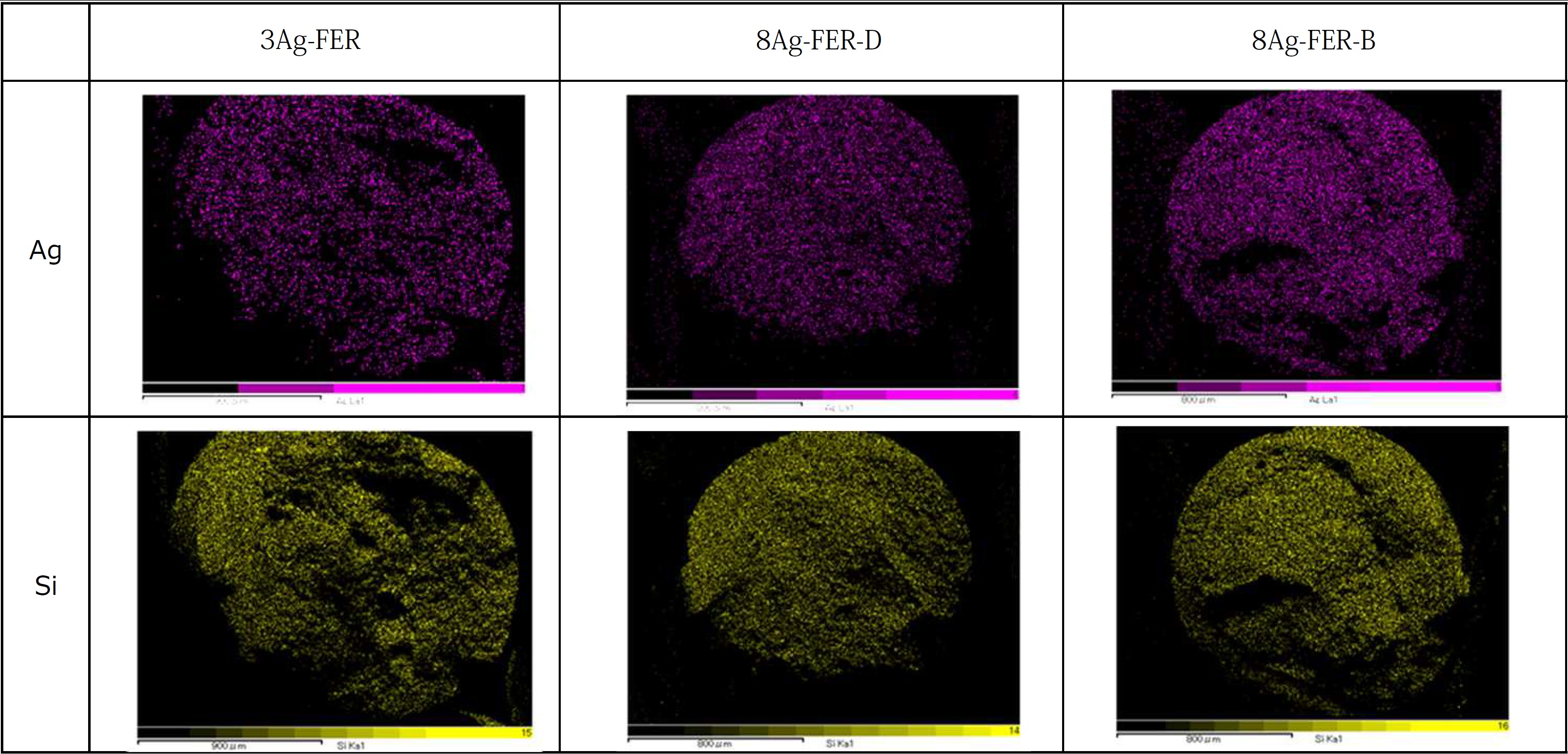}
\caption{EDX measurement results of cross section of Ag-zeolite pellets. The upper panel shows
the distribution of silver and the lower panel shows the distribution of silicon.}
\label{fig_zeo2}
\end{figure}

Figure~\ref{fig_zeo1} shows a photograph of the fabricated Ag-zeolite pellets (8Ag-FER-D).
To confirm the distribution of silver inside the pellets, the pellet cross section of
Ag-zeolite was analyzed by energy dispersive X-ray fluorescence analysis (EDX).
Figure~\ref{fig_zeo2} shows the results.
In all of the prototypes, silver was found to be introduced to the center of the pellet.

\section{Evaluation system for radon adsorption performance}

\begin{figure}[hbt]
\centering\includegraphics[width=4.5in]{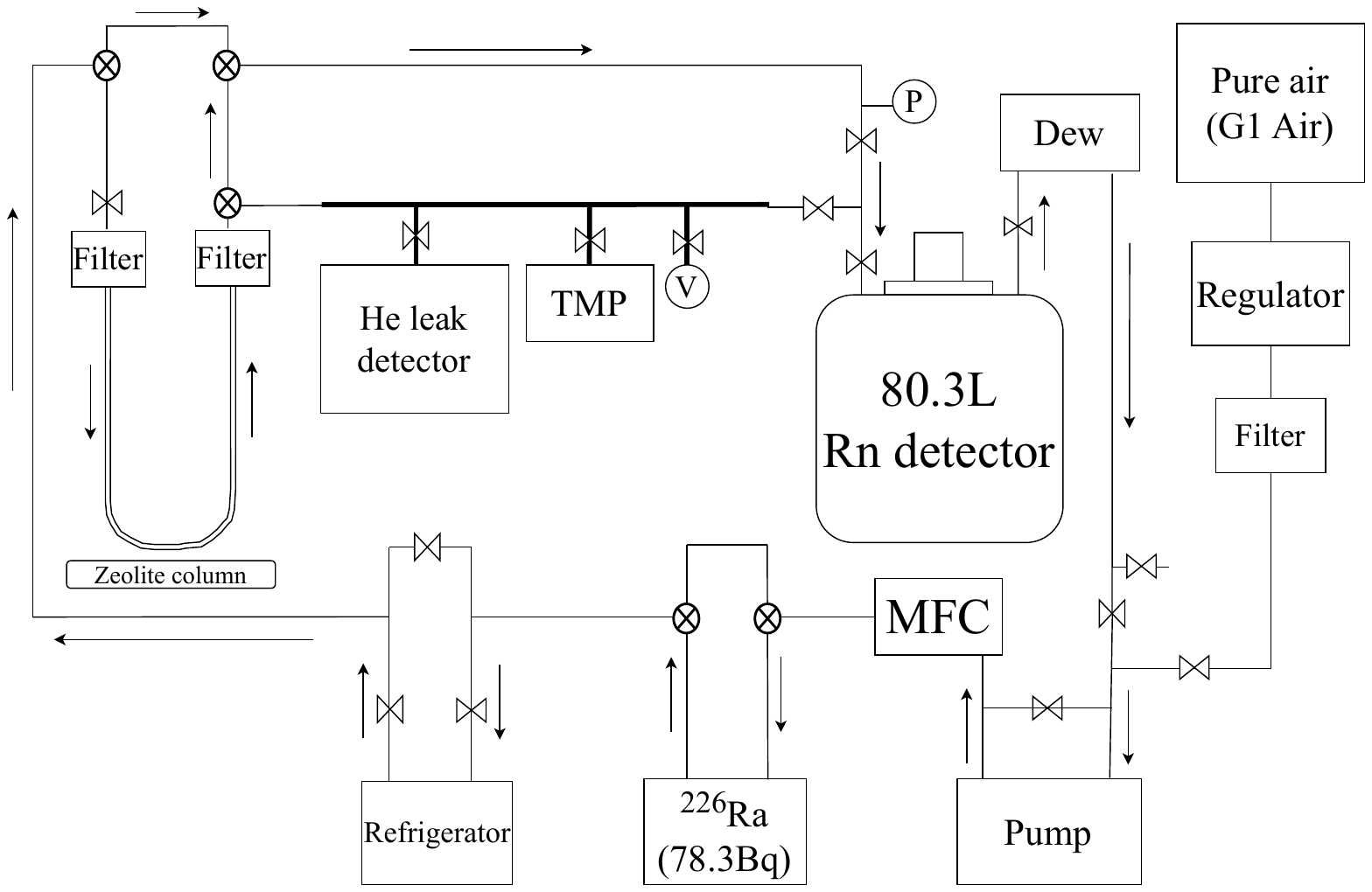}
\caption{ Experimental setup of the radon adsorption measurement system at Kobe University.
P is a pressure gauge, V is a vacuum gauge, and Dew is a dew point gauge.}
\label{fig_system1}
\end{figure}

\begin{figure}[hbt]
\centering\includegraphics[width=4in]{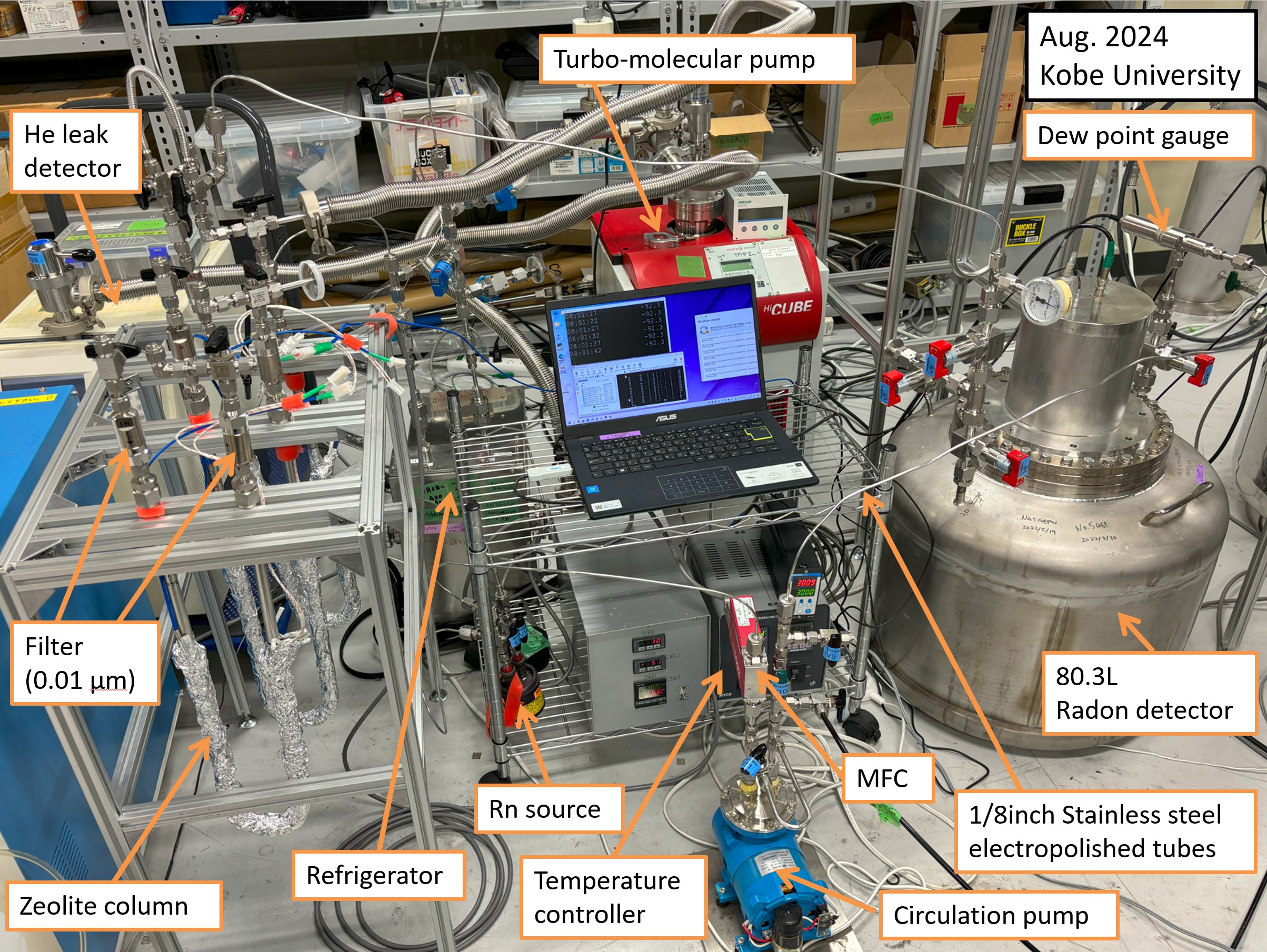}
\caption{Photo of the experimental setup at Kobe University.}
\label{fig_system2}
\end{figure}

Figures~\ref{fig_system1} and ~\ref{fig_system2} show the radon adsorption performance
evaluation system developed at Kobe University.

\begin{figure}[hbt]
\centering\includegraphics[height=3in]{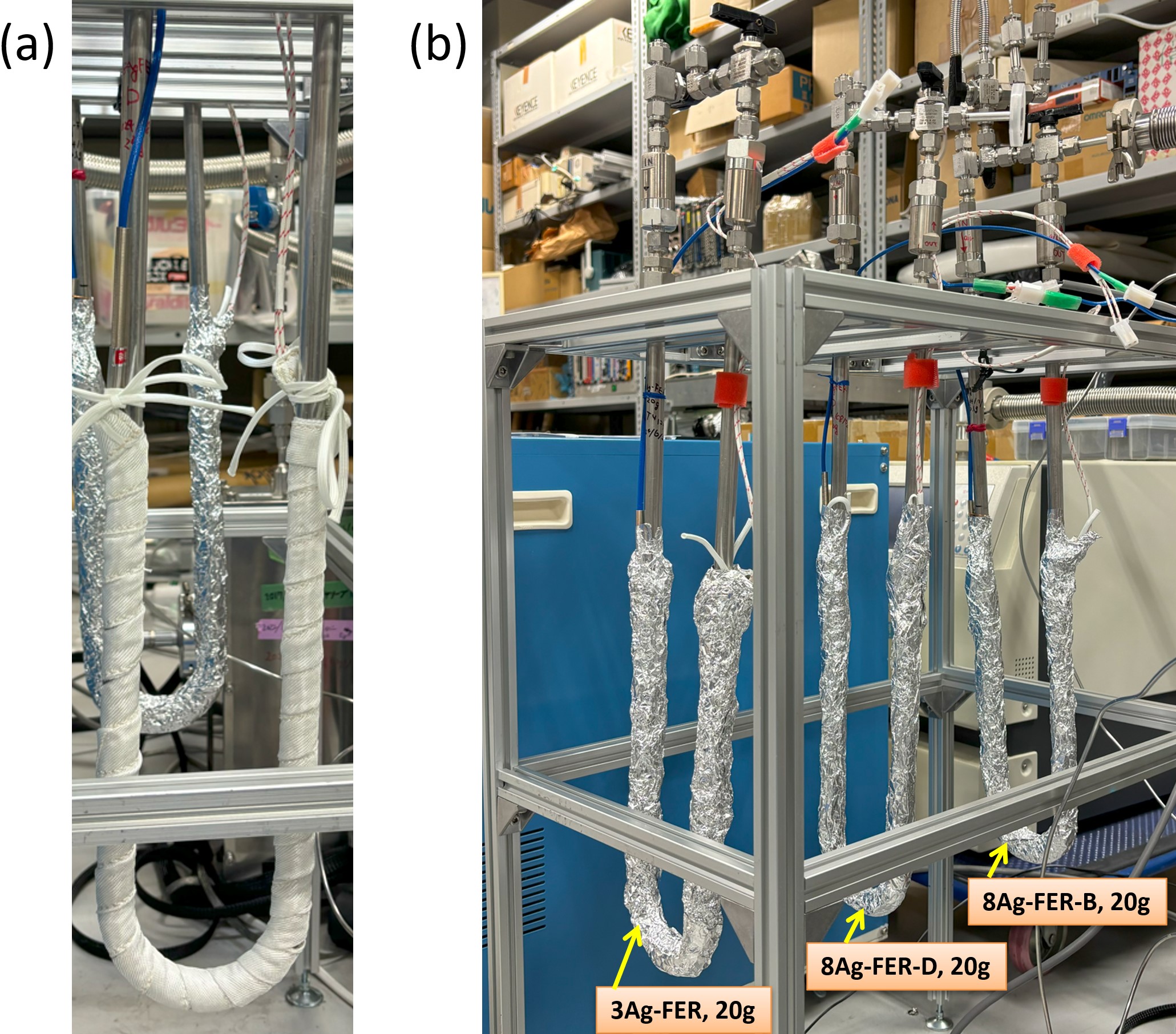}
\caption{(a) Typical, U-shaped column with a thermocouple thermometer and 150 W
ribbon heater attached. (b) In addition, columns with three types of Ag-zeolite insulated with aluminum foil.}
\label{fig_system3}
\end{figure}

Various measurements were performed using this apparatus under a room temperature environment
of about 21--25${}^\circ$C.
The apparatus was filled with pure air (TAIYO NIPPON SANSO G1 Air) at about atmospheric
pressure (differential pressure +0.003 to +0.004 MPaG) as a carrier gas.
A mass flow controller (MFC, Horiba SEC-Z500X) and a circulation pump (Enomoto Micro Pump MX-808ST-S)
were used to circulate the carrier gas
through the system at a flow rate of 1.57 to 3.00 L/min.
The temperature of the refrigerator (Taisho TC0147) was set at $-70{}^\circ$C to $-90{}^\circ$C
to control the humidity in the system.
Humidity in the carrier gas was monitored with a dew point gauge (Vaisala DMT152).
A radon source (PYLON RNC, 78.3 Bq at radiative equilibrium) was connected at all times, and
the radon concentration in the system was measured with an 80.3 L radon detector
with almost the same performance as the 80 L radon detector developed in
Refs.~\cite{80Ldet1,80Ldet2,Nakano:2017rsy}.
Ag-zeolites were placed in three nearly identical 1/2-inch stainless steel,
1-m-long U-shaped columns.
Figure~\ref{fig_system3} shows a photograph of the U-shaped columns used for
the measurement using a 20 g sample.

A typical operating procedure is as follows.
\begin{enumerate}
\item Switched the U-shaped column to bypass, circulated the carrier gas with the radon
 source connected, and checked the radiative equilibrium curve of radon concentration.
\item If switching to the measurement of a different sample, the U-shaped column was
 replaced and the Ag-zeolite was baked while evacuating with a turbo molecular pump (TMP).
 The baking temperature was set at 200${}^\circ$C, typically for several hours, and was
 conducted to an vacuum level of $1 \times 10^{-1}$ to $1 \times 10^{-3}$ Pa 
 at 200${}^\circ$C.
 Evacuation was continued after the heater was turned off, and finally the vacuum in the column
 reached  $4  \times 10^{-4} \sim 8 \times 10^{-4}$ Pa at room temperature.
 The temperature setting of the refrigerator was also changed at this timing if necessary.
\item After the U-shaped column reached room temperature, switched from bypass to
 via U-shaped column and began radon adsorption measurement.
\item When the radon concentration became stable, the radon concentration ratio
 ($Rn_{ratio}$, see below) was measured. Then, the U-shaped column was baked at 200${}^\circ$C
 while the carrier gas circulation through the U-shaped column was continued
 to remove the radon adsorbed in the column.
 After this, returned to step-(1).
\end{enumerate}

\section{Estimation of retention time and radon adsorption coefficient}

The time that radon stays in the adsorbent is generally referred to as retention time.
In Refs.~\cite{zeo1,zeo2}, the performance of radon adsorption is evaluated using
the radon adsorption coefficient ($K$ [m$^3$/kg]), which is defined as 
\begin{equation}
 K = \frac{F \times RT}{m}, 
\end{equation}
where
$F$ [m$^3$/day] is air flow rate,
$RT$ [day] is radon retention time on adsorbent, and
$m$ [kg] is mass of adsorbent.
In this study, $RT$ and $K$ were measured and evaluated as follows.

Since the radon source is always connected, 78.3 Bq of radon is distributed
in the system when radiative equilibrium.
The radon distribution is assumed to be proportional to the fraction
of time when radon stays in each device\footnote{Past measurements have confirmed that this assumption is valid to some extent.}. The devices with the longest radon stay time
in this system are the U-shaped column and the radon detector, and the other devices
such as piping are ignored. When the flow rate of the carrier gas is $F$ [m$^3$/day],
the time that radon stays in the radon detector ($T$ [day]) is $T = 0.0803 / F$,
given the volume of the radon detector (0.0803 m$^3$).
Also, radon should stay in the U-shaped column for the retention time ($RT$ [day]).
Therefore, the fraction of radon that stays in the radon detector during circulation
through the U-shaped column (called radon concentration ratio: $Rn_{ratio}$)
can be expressed as
\begin{equation}
 Rn_{ratio} = \frac{T}{RT+T} = \frac{0.0803}{F \times RT + 0.0803}. 
\end{equation}

On the other hand, the measured value of $Rn_{ratio}$ can be obtained from the ratio of
the radon concentration when circulating via the U-shaped column to that 
when the U-shaped column is bypassed, like 
\begin{equation}
 Rn_{ratio} = \frac{\rm Rn \ concentration \ via \ column}{\rm Rn \ concentration \ bypassing \ column}.
\end{equation}
When calculating $Rn_{ratio}$ from a measurement, the radiative equilibrium curve is used as the denominator in Eq.~(3).
Therefore, in this study, the measured radon concentration ratio was obtained
when the radon concentration was stable, then $RT$ was obtained from Eq.~(2),
and then $K$ was evaluated from Eq.~(1).

Possible systematic error in this evaluation method is gas flow rate uncertainty (1\%).
This uncertainty is taken into account in the measurement results in the next section.

\section{Results}

\begin{figure}[hbt]
\centering\includegraphics[height=3.5in]{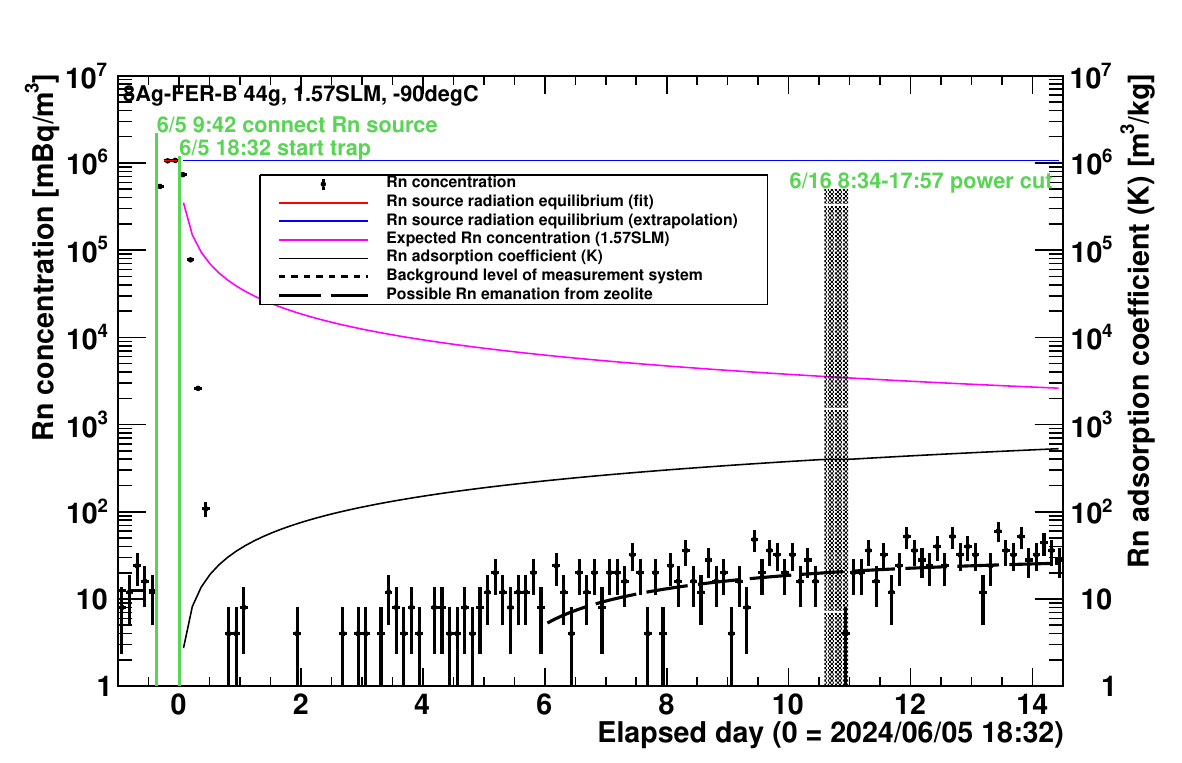}
\caption{
8Ag-FER-B 44g, 1.57 SLM, $-90{}^\circ$C (type and amount of Ag-zeolite,
carrier gas flow rate, refrigerator temperature) measurement.
Horizontal axis is elapsed day from the U-shaped column connection.
The left (right) vertical axis is radon concentration in the radon detector
(Rn adsorption coefficient).
The hatched area was not available to acquire data due to a planned power cut.
}
\label{fig_result1}
\end{figure}

At first, we filled as much as possible (44 g) of 8Ag-FER-B throughout the 1/2 inch diameter,
1-m long U-shaped column and performed radon adsorption measurements at a flow rate of
1.57 SLM (standard litter per minute) and a refrigerator setting temperature of
$-90{}^\circ$C.
The result is shown in Figure~\ref{fig_result1}.
After changing the gas circulation from U-shaped column bypass to via U-shaped column,
we confirmed that our prototype Ag-zeolite adsorbed radon.
Just before connecting the radon source to the system (around Elapsed day = $-0.5$ days in Figure~\ref{fig_result1}),
the background level of the measurement system was at $12.6 \pm 3.2$ mBq/m$^3$.
The radon concentration in the output air was lower than this background level up to
around 5 days after the column was connected.
After that, a gradual radon leakage was observed, however, it would take several months
to measure the retention time (until the measured radon concentration matched
the expected radon concentration).
Based on the results of this measurement, in order to evaluate the retention time,
we took measures to reduce the amount of zeolite (from 44 g to 20 g)
and increase the flow rate (from 1.57 SLM to 3.00 SLM) in subsequent measurements.
For 3Ag-FER, the humidity dependence of the radon adsorption coefficient was measured
at two different refrigerator temperature settings, $-70{}^\circ$C and $-90{}^\circ$C.

\begin{figure}[hbt]
\centering\includegraphics[height=2.6in]{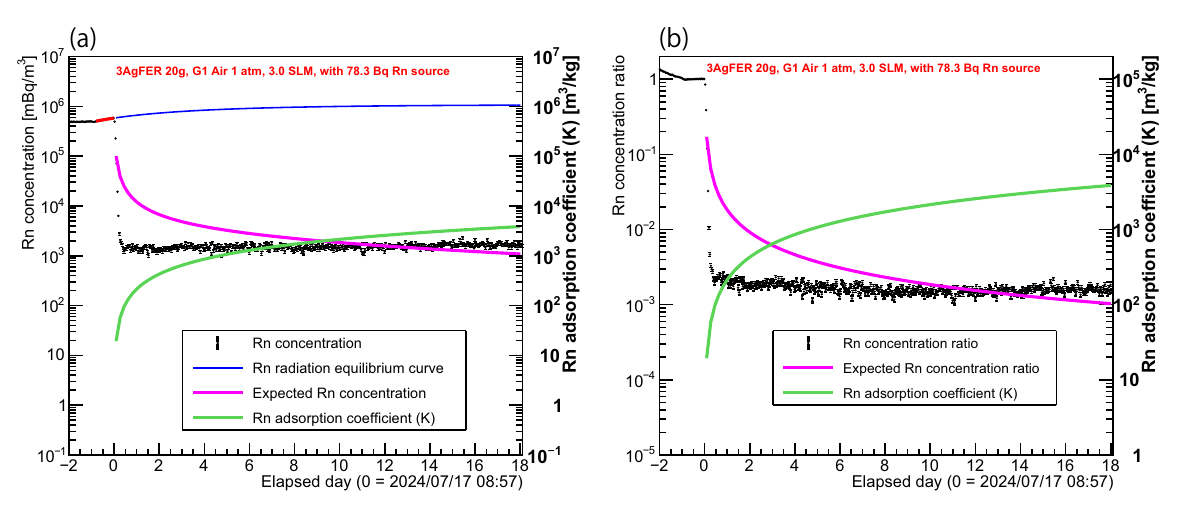}
\caption{3Ag-FER 20g, 3.00 SLM, $-90{}^\circ$C measurement.
(a) Rn concentration. (b) Rn concentration ratio.}
\label{fig_result2}
\end{figure}

Figure~\ref{fig_result2} shows an example of a radon adsorption test after
the conditions were changed.
Figure~\ref{fig_result2}(a) is a plot of the same definition as in Figure~\ref{fig_result1}.
Reducing the amount of adsorbent and increasing the flow rate enabled stable radon
concentration measurements, even via the U-shaped zeolite column.

The data points in Figure~\ref{fig_result2}(b) show the radon concentration ratio from Eq.~(3).
They were obtained by dividing the measured radon concentration (black dots in Figure~\ref{fig_result2}(a))
by the radiative equilibrium curve with radon source (blue line in Figure~\ref{fig_result2}(a)).
The average value of the data points of the radon concentration ratio in Figure~\ref{fig_result2}(b)
in the stable region (after Elapsed day = $8$ days) is the measured value of the radon concentration ratio.
The purple line in Figure~\ref{fig_result2}(b) is the value of the radon concentration ratio
obtained by Eq.~(2) when the number of elapsed days = retention time, and the elapsed time at
the intersection of this curve and the measured value of the radon concentration ratio is
the measured value of $RT$.
When the elapsed day = retention time, the green line in Figure~\ref{fig_result2}(a) and (b) shows the $K$ obtained by Eq.~(1),
and the measured value of $K$ is obtained from the measured value of $RT$ with this function.
In this measurement case, we obtained $Rn_{ratio} = 0.0016$, $RT = 12$ days, and $K = 2.6 \times 10^3$ m$^3$/kg, respectively.

\begin{figure}[hbt]
\centering\includegraphics[height=4.0in]{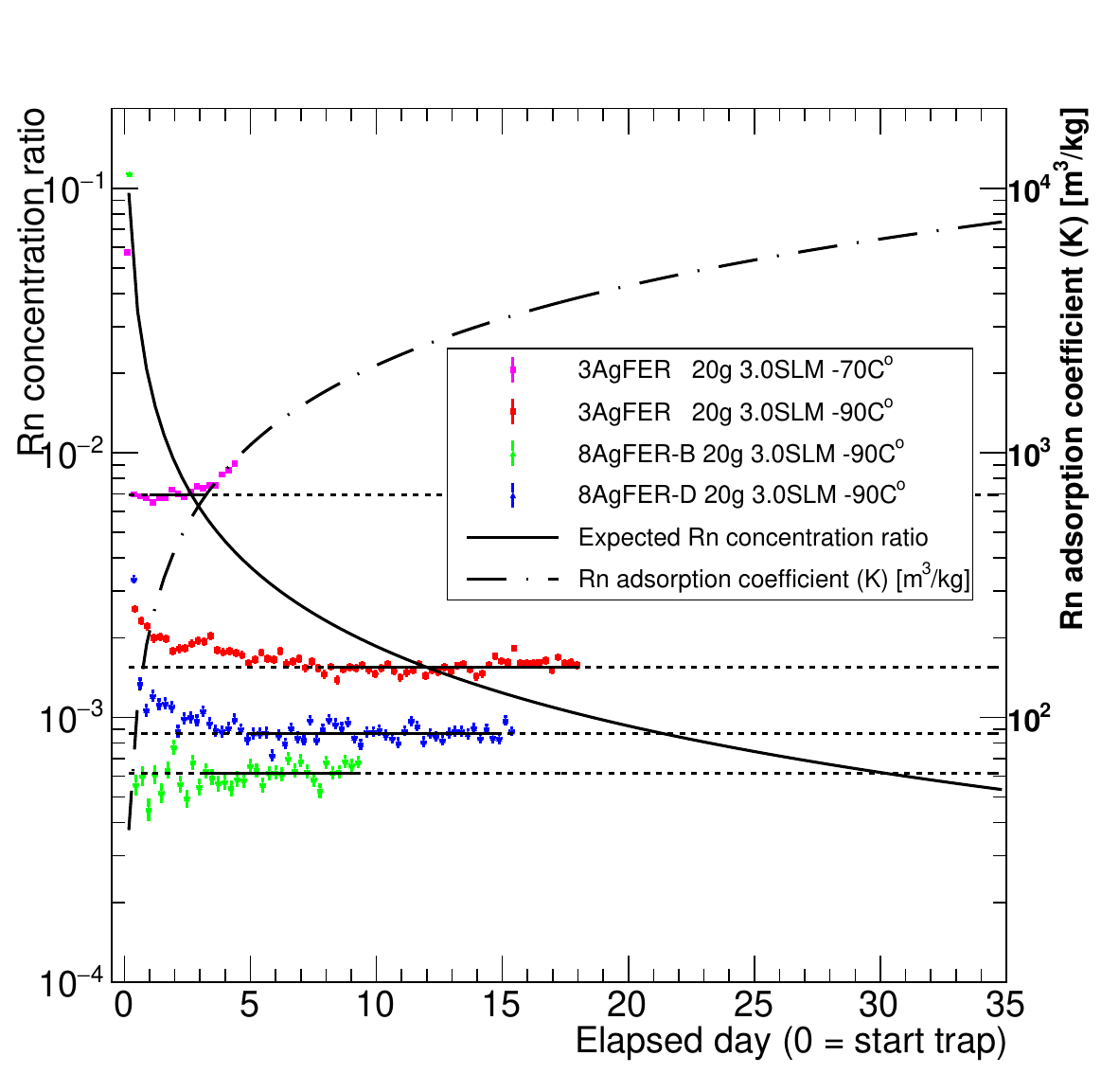}
\caption{Measurement results using 20 g of adsorbents. In each measurement data, the horizontal
solid line is the measurement period to obtain the $Rn_{ratio}$ value, and the horizontal
dotted line is the extrapolation of that line.
The elapsed day at the intersection of these horizontal lines and the black curve
of the expected radon ratio is the measured value of $RT$, and the dotted-dashed line
at that elapsed day shows the measured value of $K$.}
\label{fig_result3}
\end{figure}

Figure~\ref{fig_result3}  summarizes the results of each measurement in this study
where the mass of the adsorbent material was matched to 20 g.
The $Rn_{ratio}$ was obtained for each measurement. Periods when $Rn_{ratio}$  was not stable,
such as immediately after the start of radon trapping, were excluded from the measurement
of mean values. The values of $RT$ and $K$ obtained in this study and those in preceding
studies are summarized in Table~\ref{table_result}.
\begin{table}[hbt]
\caption{Summary of retention time ($RT$) and radon adsorption coefficient ($K$). The value with (*) are estimated by us.}
\label{table_result}
\centering
\begin{tabular}{lcccc}  \toprule
Sample    & Refrigerator    & Adsorbent  & Retention time  & Radon adsorption \\
          & setting         & mass [kg]  & [day]           & coefficient [m$^3$/kg] \\ \midrule
\multicolumn{5}{c}{(This work)} \\
3Ag-FER   & $-70{}^\circ$C  & 0.020      & $ 2.65 \pm 0.04$ & $ 573 \pm  4$   \\
3Ag-FER   & $-90{}^\circ$C  & 0.020      & $11.97 \pm 0.14$ & $2563 \pm 14$   \\
8Ag-FER-B & $-90{}^\circ$C  & 0.020      & $30.12 \pm 0.48$ & $6506 \pm 81$   \\
8Ag-FER-D & $-90{}^\circ$C  & 0.020      & $21.38 \pm 0.27$ & $4618 \pm 35$   \\
\multicolumn{5}{c}{(Preceding studies)} \\
Ag-ZSM-5~\cite{zeo1}  &     & $8.85 \times 10^{-3}$  & 10.69 & 3500   \\
Ag-ETS-10~\cite{zeo1} &     & $16.3 \times 10^{-3}$  & 19.16 & 3400   \\
Ag-ETS-10($18{}^\circ$C)~\cite{zeo2} &    & $2 \times 10^{-4}$ &  & $(140 \pm 28) \times 10$   \\
Activated Carbon      &     & \multirow{2}{*}{4.65}     & \multirow{2}{*}{1.9}      & \multirow{2}{*}{6.47 (*)} \\
Fiber~\cite{Takagi}   &     &                           &                           &  \\
Granular activated    &     & \multirow{2}{*}{26.9}     & \multirow{2}{*}{10.16}    & \multirow{2}{*}{4.96 (*)} \\
carbon~\cite{Takagi}  &     &                           &                           &  \\
Cooled activated      &     & \multirow{2}{*}{18.8 (*)} & \multirow{2}{*}{16.5 (*)} & \multirow{2}{*}{379 (*)} \\
carbon($-60{}^\circ$C)~\cite{SK-radon,Nakano:2017rsy} &  &           &              &  \\
\bottomrule
\end{tabular}
\end{table}
Interpretations on these data are given in the next section.

\section{Discussion}

In this section, the obtained results are discussed.

\subsection{Radon emanation from Ag-FER}

The dashed line in Figure~\ref{fig_result1} is a fit with the radiative equilibrium function of
radon assuming radon emanation in the measurement system
containing Ag-zeolite after Elapsed day = 6,
yielding the radon concentration at radiative equilibrium to be $31.4 \pm 2.9$ mBq/m$^3$.
Subtracting the background level of $12.6 \pm 3.2$ mBq/m$^3$ from the measurement system,
the radon emanation becomes $18.8 \pm 4.3$ mBq/m$^3$. 
Assuming that all of these are emanated from the Ag-zeolite,
and considering the amount of the zeolite and the volume of the radon detector,
the radon emanation from 8Ag-FER-B becomes 45 mBq/kg at the 90\% upper limit.
It may emanate less radon than the Ag-zeolite in Ref.~\cite{zeo2}, but
may emanate more radon than the low-radioactivity zeolite in Ref.~\cite{ogawa}.
This radon emanation level is comparable to those of activated carbons,
which have been used in underground experiments~\cite{ac1,ac2}.
Since the processed air was about 1 mBq/m$^3$ for 1 to 2 days after the Ag-zeolite
columns were connected, it is considered to be sufficiently applicable to
underground experiments through operational design such as parallelization of the zeolite
columns and alternation of radon removal from the air and baking of the column.

\subsection{Dependence of radon adsorption on humidity}

3Ag-FER was measured for radon adsorption under two different humidity conditions
in the carrier gas at refrigerator settings of $-70{}^\circ$C and $-90{}^\circ$C.
The typical dew point temperatures of the carrier gas in the system during column bypass
are $-66{}^\circ$C and $-86{}^\circ$C, respectively. As a result, as shown in Table~\ref{table_result},
the value of $K$ at the refrigerator setting of $-70{}^\circ$C is about 1/4.5 smaller
than the value at $-90{}^\circ$C. The decrease in radon adsorption capacity under high
humidity environment was noted in previous studies by Ref.~\cite{zeo1,zeo2}, and
this characteristic was also confirmed in this study.
The typical dew point temperature of the carrier gas in the system after the zeolite column
was connected was $-92{}^\circ$C (see Appendix A).
This indicates that using a carrier gas of lower dew point temperature than this
should prevent moisture adsorption on the Ag-zeolite, and in that case,
radon removal-only operation would be possible for a long period of time without baking.

\subsection{Dependence of radon adsorption on silver content and silver introduction status}

In this study, radon adsorption measurements were performed using Ag-zeolites
with silver introductions adjusted to 3\% and 8\%.
If the amount of silver is proportional to the amount of radon adsorbed,
then $RT$ should vary in proportion to the amount of silver in this measurement system.
Considering the actual amount of silver introduced in Table~\ref{table_zeo2}, and comparing
the measurements at the refrigerator setting of $-90{}^\circ$C in Table~\ref{table_result},
8Ag-FER-B and 3Ag-FER have a nearly proportional relationship between the amount of
silver and the $RT$ value. On the other hand, the value of $RT$ for 8Ag-FER-D
is considerably smaller than the value of $RT$ expected from 3Ag-FER.
Therefore, when introducing about 8\% silver to FER, the "-B" method is better than "-D" method.
Silver is introduced as metallic silver or silver ions, but due to the difference
in the introduction method, the proportion of metallic silver is higher in "-D" than in "-B".
This suggests that the most effective silver introduction for radon adsorption may be
silver ions rather than metallic silver.

\subsection{Radon adsorption performance and application example for underground experiments}

In this study, 8Ag-FER-B showed the best radon adsorption properties in air, with a radon
adsorption coefficient $K = 6506 \pm 81$ m$^3$/kg.
This is higher than the performance of Ag-zeolites to adsorb radon
in air at room temperature tested in previous studies in Ref.~\cite{zeo1,zeo2}.
The granular activated carbon in Table~\ref{table_result} is equivalent to that used
in the Super-Kamiokande radon removal system~\cite{Fukuda:2002uc}, and the cooled
activated carbon is the actual performance in the Super-Kamiokande radon removal system.
8Ag-FER-B was found to have a higher radon adsorption coefficient than any of these adsorbents.

We now consider the application of 8Ag-FER-B to the Hyper-Kamiokande (HK) detector~\cite{HK},
the next generation of large underground water Cherenkov detector.
The air purification system at HK is currently planing to produce dry air
with a dew point temperature of $-70{}^\circ$C at a flow rate of 54 Nm$^3$/hour
to supply the water purification system and the HK detector. 
The radon concentration in this dry air is expected to be about 50 Bq/m$^3$,
so radon must be reduced to 1/50000 to achieve a radon concentration of 1 mBq/m$^3$.
The required retention time is $RT = 59.68$ days because radon will decay and decrease
in the adsorbent while being adsorbed and retained. 
If the dew point temperature in the dry air can be reduced to $-90{}^\circ$C by some
pretreatment, about 12 kg of Ag-zeolite would be required in the case of
8Ag-FER-B to reduce the radon concentration in air to 1/50000 at a flow rate of 54 Nm$^3$/hour.
Even if two systems are constructed to operate in parallel for radon removal from the air and
baking, the total weight would be 24 kg. 
In terms of volume, 24 kg of 8Ag-FER-B is about 35 L.
The radon removal system in Super-Kamiokande~\cite{Fukuda:2002uc} uses about 8 m$^3$
of room temperature activated carbon, and so on.
Compared to this, it is likely that the system with Ag-zeolite can be made much smaller.

\section{Conclusion}

In this study, we focused on Ag-FER, which has been reported to have high performance
against xenon in air, and measured its radon adsorption properties.
The silver ion-exchanged zeolites used in these measurements were based on the zeolite
from Tosoh Corporation, and silver ions were introduced into the zeolite by Sinanen Zeomic Co., Ltd.
A dedicated measurement system was constructed at Kobe University, and five types of
measurements were systematically performed using U-shaped columns of almost the same shape,
yielding the following results.

As in the preceding studies, radon adsorption in air at room temperature was confirmed.
The intrinsic radon emanation from 8Ag-FER-B was evaluated to be less than 45 mBq/kg (90\% C.L.).
This intrinsic radon emanation level is comparable to those of activated carbons,
which have been used in underground experiments.
In the case of 8Ag-FER-B, radon removal down to 1 mBq/m$^3$ was observed for about 1 to 2 days
after the start of the radon adsorption test, which indicates that 8Ag-FER-B can be applied
to underground experiments. On the other hand, radon adsorption efficiency was found to
be sensitive with respect to air humidity.
In order to increase the efficiency of radon adsorption, dehumidification to a dew point
temperature of about $-86{}^\circ$C or lower is desirable.
To increase the efficiency of radon adsorption, it was effective to increase the amount
of silver. Furthermore, it was found that introducing silver in the form of silver ions,
rather than metallic silver, is likely to be a good method to increase the efficiency
of radon adsorption. The prototype of 8Ag-FER-B in this study showed higher radon
adsorption performance than any of the preceding studies, and when 8Ag-FER-B
is used in the Hyper-Kamiokande experiment, it is expected to be much smaller (12-24 kg, 17-35 L in volume)
than the existing radon removal system using room-temperature activated carbon.

\section*{Acknowledgment}

The authors would like to thank Kotaro Nagatsu (QST) and Jun Ichinose (QST) for
their helpful discussions on radon removal performance of silver-ion exchanged zeolite from air.
This work is supported by Japan Society for the Promotion of Science (JSPS) KAKENHI Grant Number 24H02243.

\bibliographystyle{ptephy}
\bibliography{main}

\begin{thebibliography}{10}

\bibitem{SK-radon}
Y.~Takeuchi et~al., Phys. Lett. B, {\bf 452}, 418--424 (1999).

\bibitem{SK-radon2}
G.~Pronost et~al., Prog. Theor. Exp. Phys., {\bf 2018}, 093H01 (2018).

\bibitem{XENON1}
E.~Aprile et~al., Eur. Phys. J. C, {\bf 82}, 599 (2022).

\bibitem{Fukuda:2002uc}
Y.~Fukuda et~al., Nucl. Instrum. Meth. A, {\bf 501}, 418--462 (2003).

\bibitem{Nakano:2017rsy}
Y.~Nakano, H.~Sekiya, S.~Tasaka, Y.~Takeuchi, R.~A. Wendell, M.~Matsubara, and
  M.~Nakahata, Nucl. Instrum. Meth. A, {\bf 867}, 108--114 (2017).

\bibitem{XENON2}
M.~Murra et~al., Eur. Phys. J. C, {\bf 82}, 1104 (2022).

\bibitem{Fukui}
M.~Fukui et~al.,
\newblock Xe adsorption performance of ag-loaded zeolite,
\newblock In {\em The 36th Zeolite Research Presentation (on-line)}. The
  Zeolite Institute of Japan (2020).

\bibitem{zeo1}
S.~Heinitz et~al., Sci Rep, {\bf 13}, 6811 (2023).

\bibitem{zeo2}
O.~Veselska et~al., Prog. Theor. Exp. Phys., {\bf 2024}, 023C01 (2024).

\bibitem{ogawa2}
H.~Ogawa et~al.,
\newblock Removal of radioactive noble gas radon from air by {A}g-zeolite
  (2024),  {{arXiv:2411.01722}}.

\bibitem{zeo3}
K.~Coopersmith et~al., J. Phys. Chem. C, {\bf 127}, 1598--1606 (2023).

\bibitem{ac1}
K.~Abe et~al., Nucl. Instrum. Meth. A, {\bf 661}, 50--57 (2012).

\bibitem{ac2}
Y.~Nakano et~al., Prog. Theor. Exp. Phys., {\bf 2020}, 113H01 (2020).

\bibitem{ogawa}
H.~Ogawa et~al., JINST, {\bf 19}, P02004 (2024).

\bibitem{80Ldet1}
K.~Hosokawa et~al., J. Phys.: Conf. Ser., {\bf 469}, 012007 (2013).

\bibitem{80Ldet2}
K.~Hosokawa, A.~Murata, Y.~Nakano, Y.~Onishi, H.~Sekiya, Y.~Takeuchi, and
  S.~Tasaka, Prog. Theor. Exp. Phys., {\bf 2015}, 033H01 (2015).

\bibitem{Takagi}
Y.~Takagi,
\newblock Measurement of radon removal performance of ambient temperature
  activated carbon in air for {H}yper-{K}amiokande (March 2024),
\newblock Available at
  \url{https://ppwww.phys.sci.kobe-u.ac.jp/seminar/pdf/Takagi_Mthesis.pdf}.

\bibitem{HK}
K.~Abe et~al.,
\newblock Hyper-{K}amiokande design report (2018),  {{arXiv:1805.04163}}.

\end{thebibliography}

\vspace{0.2cm}
\noindent

\newpage
\appendix
\section{Measured data}

Here we present data on the measured radon concentration and dew point temperature changes.

\begin{figure}[hbt]
\centering\includegraphics[height=2.5in]{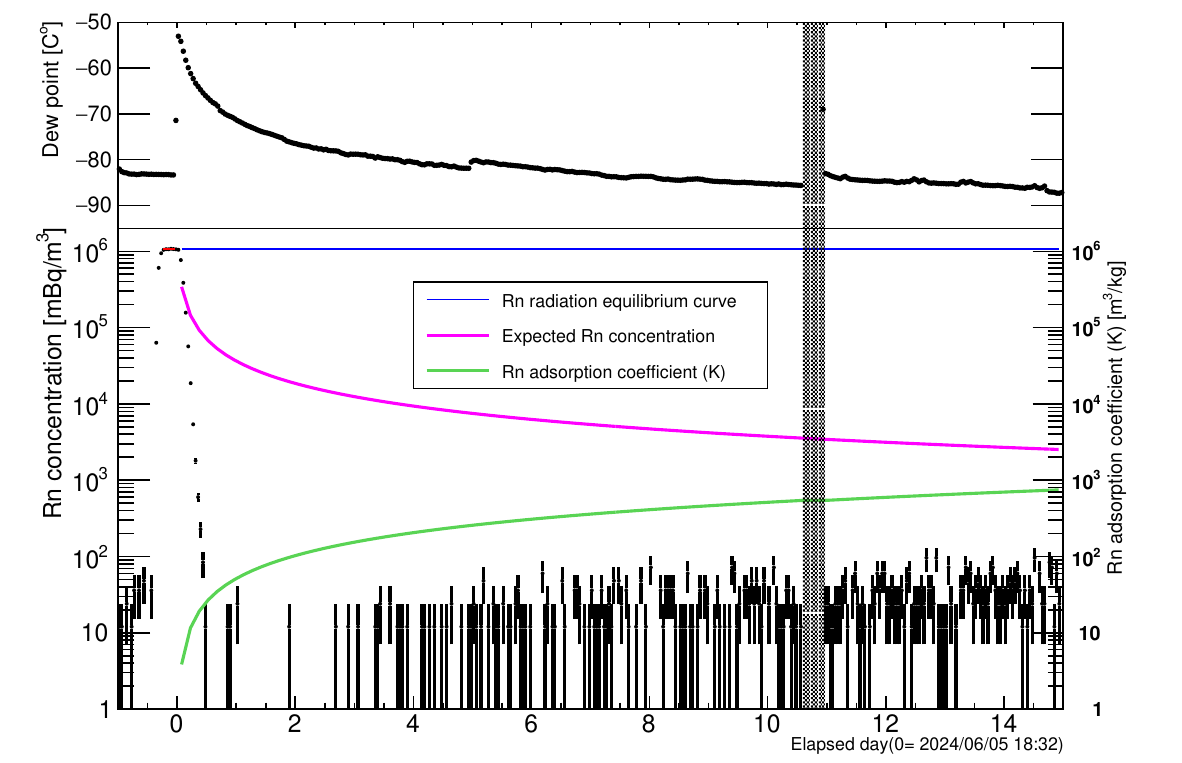}
\caption{
8Ag-FER-B 44g, 1.57 SLM, $-90{}^\circ$C (type and amount of Ag-zeolite,
carrier gas flow rate, refrigerator temperature) measurement.
Bottom panel: Variation of the radon concentration.
Horizontal axis is elapsed day from the U-shaped column connection.
The left (right) vertical axis is radon concentration in the radon detector (Rn adsorption coefficient).
The red line shows the fitting result with the Rn source radiative equilibrium function.
The blue line shows the extrapolation of the red line.
Top panel: Variation of the dew point temperature. The horizontal axis is common with the bottom plot.
The hatched area was not available to acquire data due to a planned power cut.
}
\label{fig_a1}
\end{figure}

\begin{figure}[hbt]
\centering\includegraphics[height=2.3in]{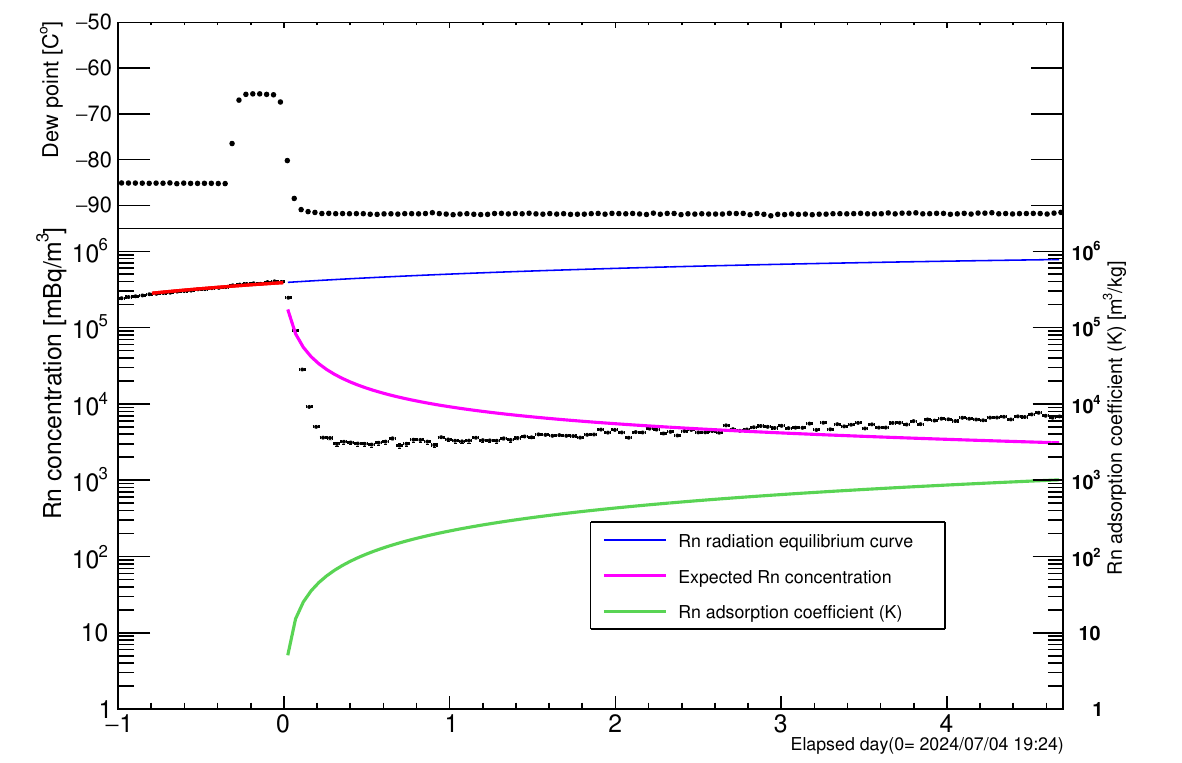}
\caption{3Ag-FER 20g, 3.00 SLM, $-70{}^\circ$C measurement. Others are same as Fig.~\ref{fig_a1}.}
\label{fig_a2}
\end{figure}

\begin{figure}[hbt]
\centering\includegraphics[height=2.3in]{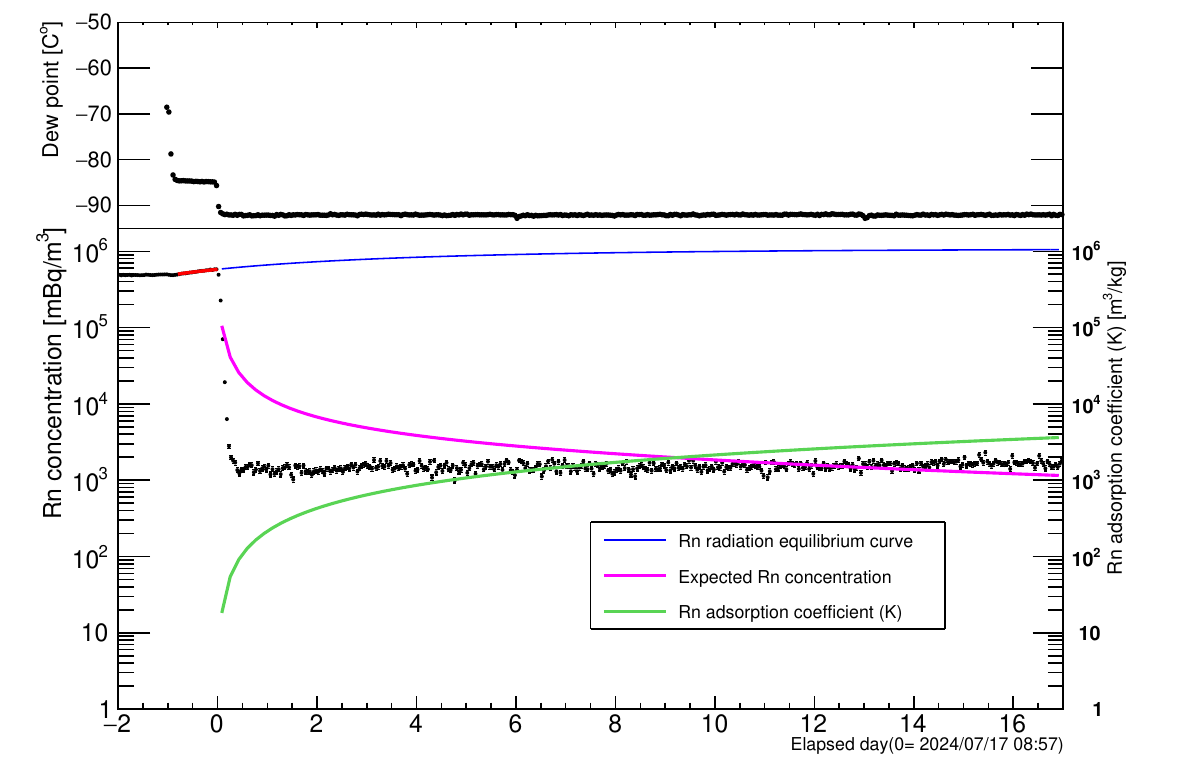}
\caption{3Ag-FER-B 20g, 3.00 SLM, $-90{}^\circ$C measurement. Others are same as Fig.~\ref{fig_a1}.}
\label{fig_a3}
\end{figure}

\begin{figure}[hbt]
\centering\includegraphics[height=2.3in]{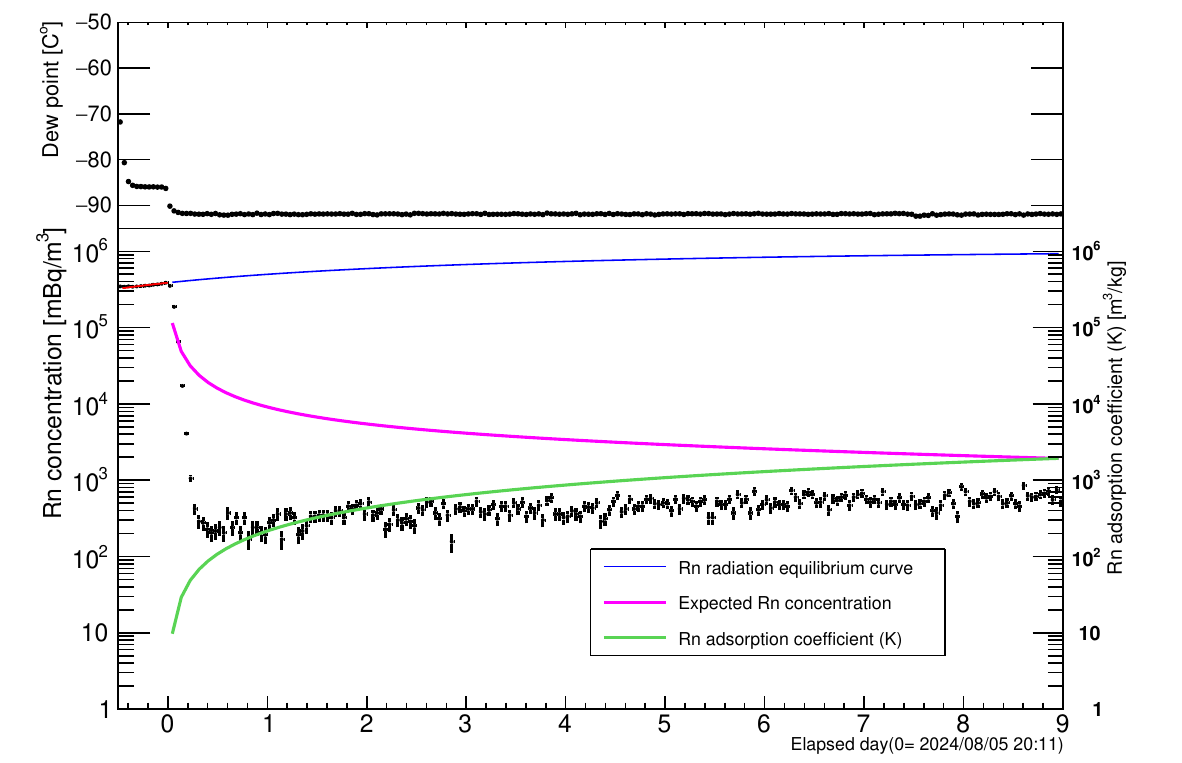}
\caption{8Ag-FER-B 20g, 3.00 SLM, $-90{}^\circ$C measurement. Others are same as Fig.~\ref{fig_a1}.}
\label{fig_a4}
\end{figure}

\begin{figure}[hbt]
\centering\includegraphics[height=2.3in]{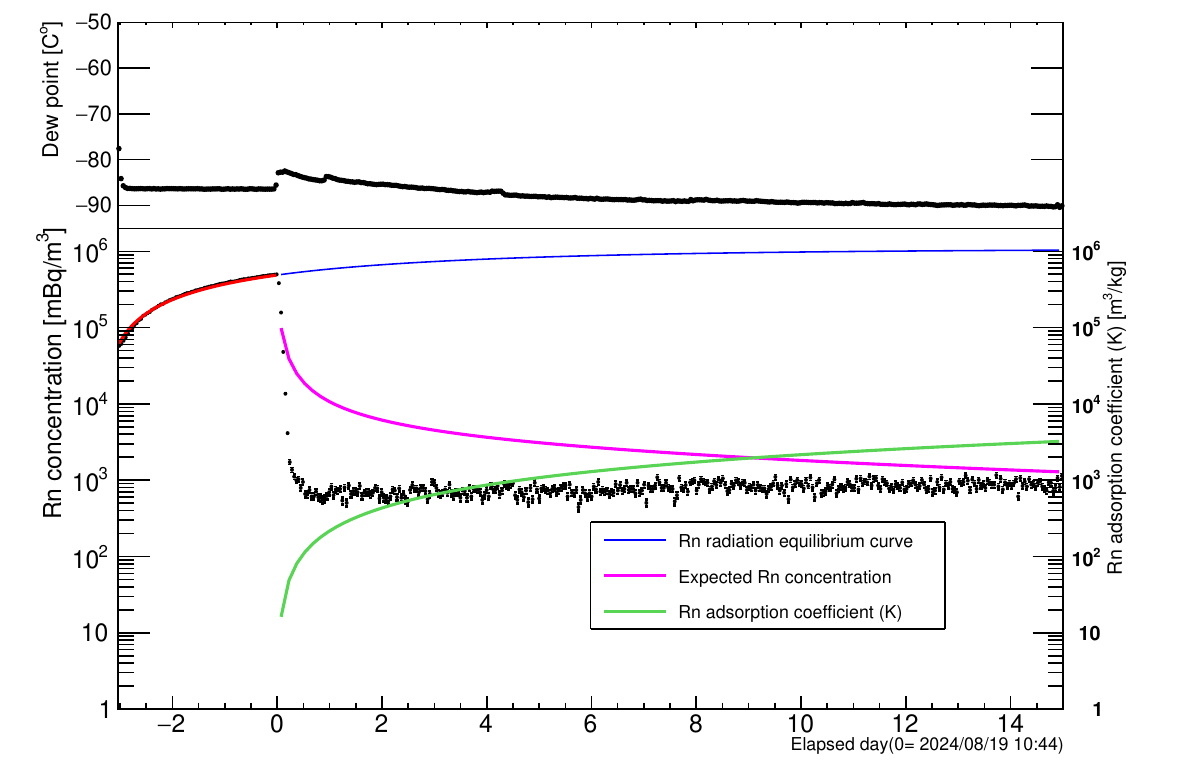}
\caption{8Ag-FER-D 20g, 3.00 SLM, $-90{}^\circ$C measurement. Others are same as Fig.~\ref{fig_a1}.}
\label{fig_a5}
\end{figure}

\end{document}